# Accelerated Patient-specific Non-Cartesian MRI Reconstruction using Implicit Neural Representations

Di Xu[1], Hengjie Liu[2], Xin Miao[3], Daniel O'Connor[4], Jessica E. Scholey[1], Wensha Yang[1], Mary Feng[1], Michael Ohliger[5], Hui Lin[1], Dan Ruan[2], Yang Yang[5] and Ke Sheng[1, *]

1 Radiation Oncology, University of California, San Francisco, 505 Parnassus Ave, San Francisco, CA 94143

2 Radiation Oncology, University of California, Los Angeles, 200 Medical Plaza, Los Angeles, CA 90095

3 Radiology at Children's Hospital Los Angeles, Keck School of Medicine, University of Southern California, 1975 Zonal Ave, Los Angeles, CA 90033

4 Collage of Arts and Sciences, University of San Francisco, 2130 Fulton Street, San Francisco, CA 94117

5 Radiology, University of California, San Francisco, 505 Parnassus Ave, San Francisco, CA 94143

Corresponding author: ke.sheng@ucsf.edu

# Abstract

The scanning time for a fully sampled MRI can be undesirably lengthy. Compressed sensing has been developed to minimize image artifacts in accelerated scans, but the required iterative reconstruction is computationally complex and difficult to generalize on new cases. Image-domain-based deep learning methods (e.g., convolutional neural networks) emerged as a faster alternative but face challenges in modeling continuous k-space, a problem amplified with non-Cartesian sampling commonly used in accelerated acquisition. In comparison, implicit neural representations can model continuous signals in the frequency domain and thus are compatible with arbitrary k-space sampling patterns. The current study develops a novel generative-adversarially trained implicit neural representations (k-GINR) for de novo undersampled non-Cartesian k-space reconstruction. k-GINR consists of two stages: 1) supervised training on an existing patient cohort; 2) self-supervised patient-specific optimization. In stage 1, the network is trained with the generative-adversarial network on diverse patients of the same anatomical region supervised by fully sampled acquisition. In stage 2, undersampled k-space data of individual patients is used to tailor the prior-embedded network for patient-specific optimization. The UCSF StarVIBE T1-weighted liver dataset was evaluated on the proposed framework. k-GINR is compared with an image-domain deep learning method, Deep Cascade CNN, and a compressed sensing method. k-GINR consistently outperformed the baselines with a larger performance advantage observed at very high accelerations (e.g., 20 times). k-GINR offers great value for direct non-Cartesian k-space reconstruction for new incoming patients across a wide range of accelerations liver anatomy.

## 1. Introduction

Magnetic resonance imaging (MRI) is a powerful, non-invasive medical imaging modality for diagnosing soft-tissue anomalies. However, the scanning time for fully sampled high-quality images can be undesirably long due to the sequential acquisition nature of k-space data and the inherent sensitivity of MRI to motion artifacts. The prolonged MRI acquisition time substantially increases the associated operational cost and restricts its applicability in clinical scenarios requiring rapid imaging. Efforts to accelerate MRI acquisition generally involve two approaches: 1) parallel imaging [1], which simultaneously acquires numerous views with multiple receiver coils, and 2) sparse sampling [2], which acquires fewer samples or utilizes more efficient sampling trajectories (e.g., radial and spiral sampling). While combining these approaches promises significantly faster scan times, accurately reconstructing images from aggressively undersampled multi-coil data remains an open challenge.

image reconstruction based on sparse samples is an ill-posed inverse problem where the solution cannot be uniquely determined. Various techniques have been developed to address the challenge over the past few decades. Among them, compressed sensing (CS) [3] was widely adopted to solve a constrained optimization problem, where regularization priors, such as total variation [4], low-rank [5], and dictionary learning [6], [7], were applied to preserve pertinent anatomical information while mitigating artifacts and noise. Though having achieved vast success, CS exhibits the following notable limitations. First, the assumed sparsity in the spatial or k-t domain is an approximation whose goodness varies with anatomical complexity and physiological irregularity. The approximation can lead to degraded reconstruction quality and diagnostic values. Second, CS can rapidly lose effectiveness in artifact mitigation with suboptimal sampling patterns and/or aggressive acceleration. Third, effective CS reconstruction requires case-wise fine-tuning of regularization parameters, without which the real-world performance of CS can be significantly compromised [8], [9]. Fourth, CS reconstruction can be slow due to its iterative nature. The last two challenges have significantly hampered its clinical adoption [10], [11].

Recent advances in deep learning (DL) have introduced a data-driven framework for accelerated MRI reconstruction. Unlike explicitly designed CS methods, DL leverages the extensive information in training data to learn the reconstruction representation mapping. DL not only matches or surpasses the quality of CS but also offers significantly faster reconstruction. Previous studies have investigated accelerated MRI reconstruction through the application of convolutional neural networks (CNNs), recurrent neural networks (RNN), Transformers, and hybrid models combining these network architectures [12], [13], [14], [15], [16], [17], [18]. For instance, Schlemper et al. introduce a cascaded CNN structure to progressively remove noises and artifacts in the accelerated MRI images. This deep cascade architecture improves the output by leveraging spatial information from the previous layers and empowers the model to capture complex data patterns and dependencies within MRI data [12]. Lonning et al. propose a recurrent inference machine (RIM) that iteratively optimizes undersampled MRI using trained RNN priors. The temporally recurrent design of RIM makes it particularly advantageous for dynamic MRI reconstruction, as it captures sequential dependencies across frames [14]. Guo et al. introduce ReconFormer [15], a recurrent Transformer structure built with Recurrent Pyramid Transformer Layers (RPTLs), to better capture multi-scale information and deep feature correlation. ReconFormer is efficient and lightweight while maintaining high fidelity in reconstructing accelerated MRI images. However, these algorithms operate exclusively in the image domain to refine the image representation rather than in the native frequency domain.

In accelerated MRI reconstruction, k-space or dual-domain-based algorithms have distinct advantages over image-domain-only methods for two main reasons. First, k-space processing allows the model to handle raw signals and better retain high-frequency information (fine anatomical details in the image domain), which is degraded in image-domain-based DL reconstructions [19], [20]. Second, k-space processing can incorporate MRI physics to further reduce artifacts and noise. There are a few studies investigating the feasibility of processing k-space signals using CNNs and Transformers [21], [22], [23], [24]. For instance, SPIRiT, the iterative self-consistent parallel imaging reconstruction from an arbitrary k-space approach, was formulated to use CNNs to directly process in k-space and enable the reconstruction of high-frequency image details and textures [23]. DuDoUniNeXt, a dual-domain approach combining CNNs with Vision Transformers, was proposed to operate simultaneously in both k-space and image space [24].

Nevertheless, these methods have not been used to reconstruct non-Cartesian acquisitions, e.g., radial, spiral, and koosh ball sequences that have shown superior resilience to artifacts and are widely used in accelerated acquisitions. The challenge is fundamental to convolutional filters in CNNs and Transformers that are compatible only with Cartesian data points, as illustrated in Figure 1.

Alternatively, implicit neural representations (INRs) were proposed to leverage multi-layer perceptron (MLP) and periodic activation functions for representing continuous and differentiable signals with fine details [25]. INRs have been shown to successfully solve challenging boundary value problems, such as the Helmholtz and wave equations [26], which were traditionally analyzed using Fourier Transforms [25]. Such capabilities highlight the potential of INRs for representing Fourier sequences within the framework of MRI k-space. Several pioneering studies have employed INRs to represent non-Cartesian k-space data. Specifically, Spieker et al. [27] proposed ICoNIK for reconstructing motion-resolved abdominal MRI directly in k-space with radial sampling trajectories. The framework is designed and optimized on a single MRI volume to reconstruct high-quality respiratory-resolved images. Moreover, Shen et al. [28] proposed a neural representation learning methodology with a prior embedding (NeRP) framework to reconstruct computational images from sparsely sampled measurements in frequency domains. The feasibility of reconstructing radial sampling k-space data using NeRP has been demonstrated. NeRP is designed to leverage prior knowledge from the object's historical scans while optimizing reconstruction based on the sparsely sampled measurements of the current scan. Therefore, the application of NeRP is confined to scenarios where recent scans of the same patient are available, precluding de novo scans or patients who experienced significant anatomical changes due to treatment intervention, disease progression, or other physiological changes.

One approach to address this limitation [25] is to train INRs on the common anatomical patterns of a diverse patient population, forming a generalizable prior that can be fine-tuned using newly acquired patient-specific sparse data for optimized, individualized reconstruction. In this work, we develop k-GINR, a patient-specific INR model with generative-adversarial training [29], designed for accelerated MRI reconstruction directly from undersampled non-Cartesian k-space data. Including generative-adversarial training enhances prior convergence, improving accuracy and adaptability in capturing complex anatomical details across individual patients[29]. The performance of k-GINR is evaluated using the UCSF StarVIBE T-1 weighted liver dataset. The remainder of this manuscript is organized as follows: Section 2 details the k-GINR methodology, Section 3 describes the data cohort, experimental setup, and results, and Sections 4 and 5 provide discussion and conclusions.

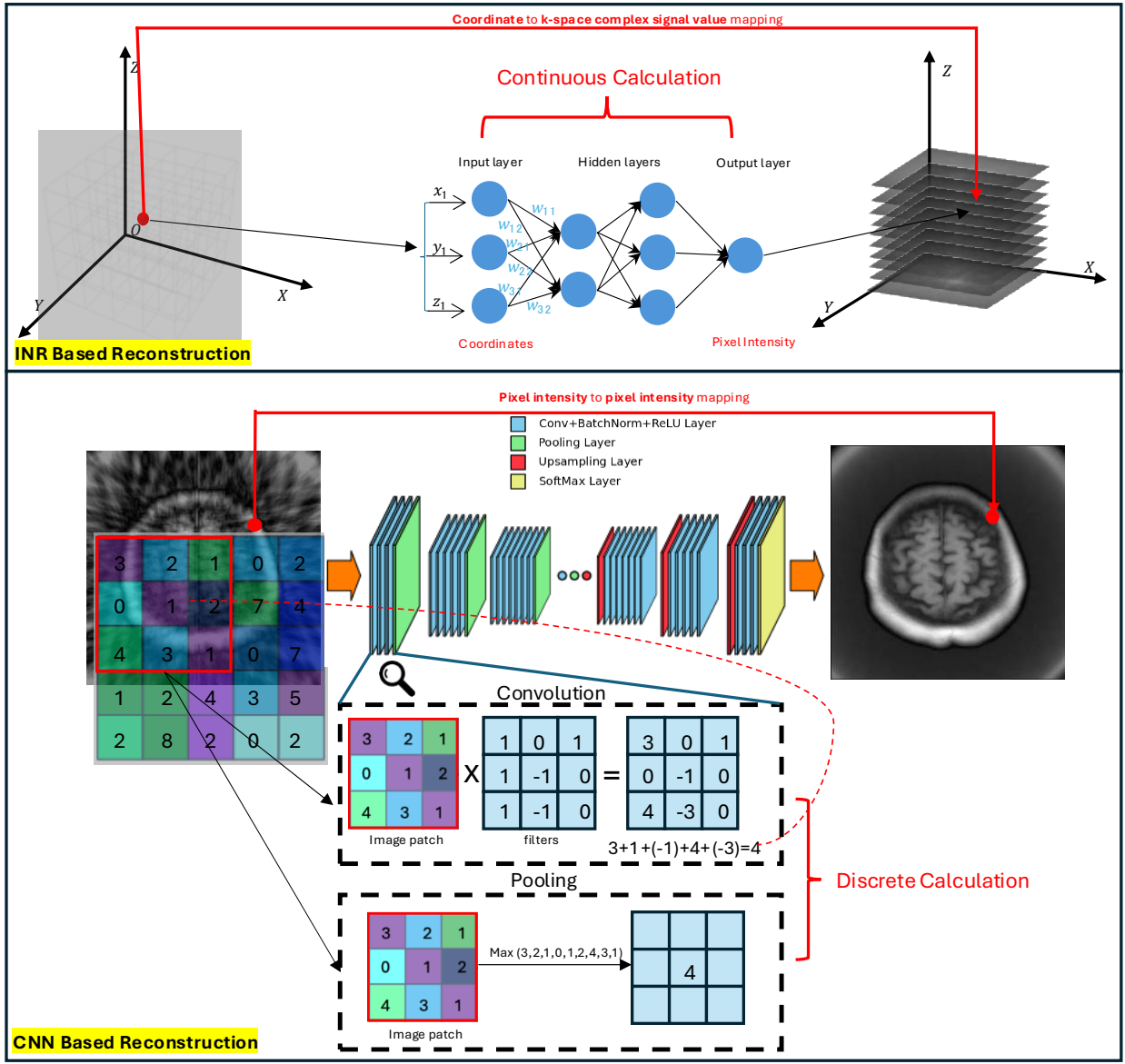

**Figure 1**: The mechanism comparison between INR (top panel) and CNN (bottom panel) based reconstruction algorithms.

## 2. Materials and Methods

As shown in Figure 2, the k-GINR pipeline consists of two stages: 1) Supervised training on prior acquisition and 2) self-supervised patient-specific reconstruction optimization. For stage 1, the INR network is trained on a diverse cohort of patients representing consistent anatomical structure with uniform fields of view, aiming to establish a training prior that encapsulates the universal features inherent to common anatomy. Fully sampled k-space acquisition is utilized for training supervision. We also incorporate a generative adversarial network (GAN) [29] to improve convergence. In stage 2, undersampled k-space data of an individual patient is used to tailor the prior-embedded INR for patient-specific optimization. $L_2$ loss is used in this stage. The entire INR operates *exclusively* in the k-space to leverage the intrinsic k-space information, thus avoiding the loss of imaging domain information. It is also essential to point out that stage 1 is fully supervised by fully sampled ground truth (GT) k-space signals, but stage 2 is self-supervised by accelerated k-space signals.

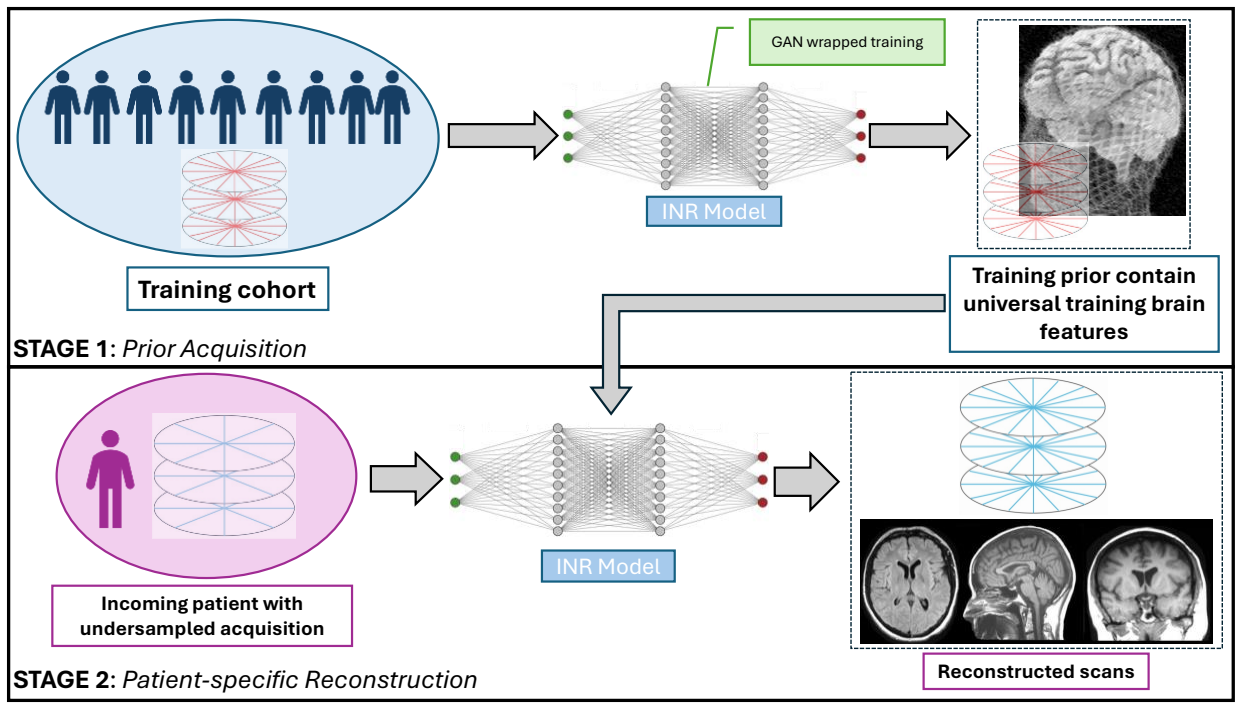

**Figure 2**: The overall framework of k-INR. Stage 1 illustrates the universal prior training process. Stage 2 illustrates the patient-specific optimization process.

## 2.1 Theoretical Difference between INR and CNN Architecture in Reconstruction

Figure 1 highlights the fundamental theoretical differences between INR- and CNN-based architectures for image reconstruction. As briefly discussed in the Introduction, INR utilizes a continuous mapping approach where spatial coordinates $(x, y, z)$ serve as inputs to a fully connected MLP with outputs of their corresponding pixel intensities. Such a paradigm represents the image as a continuous function, enabling high-fidelity reconstruction without being constrained by discrete grid structures or interpolation artifacts. In contrast, the CNN-based reconstruction framework, where noisy pixel intensities serve as inputs and reconstructed pixel intensities serve as outputs, operates on discrete pixel intensity mapping through convolutional layers with predefined filters, followed by pooling operations for feature aggregation and down-sampling. Such discrete signal processing can result in signal degradation, particularly when handling complex distribution patterns, such as non-Cartesian or significantly sparsely sampled data, as CNNs inherently assume input in a fixed grid-based structure [30]. On the other hand, the coordinate-based representation of INR offers superior flexibility. It is particularly advantageous for tasks requiring precise interpolation or the handling of nonuniform Cartesian sampling patterns [25], [31], [32]. Furthermore, the continuous nature of INR frameworks allows for more detailed reconstructions compared to the pixelated outputs often observed in CNN-based methods.

## 2.2 Problem Formulation

In the inverse MRI reconstruction problem, the forward process can be formulated as:

$$\dot{K} = \dot{M}\dot{F}\dot{S}\dot{I} + \varepsilon \quad (1)$$

Where $\dot{I}$ is the MR image of the target object, $\dot{K}$ is the sampled sensor measurements in k-space. $\dot{M}$ is the undersampling mask in k-space, $\dot{F}$ is the Fourier transform of the imaging system, $\dot{S}$ is the coil sensitivities and $\tilde{\varepsilon}$ is the acquisition noise. MRI image reconstruction aims to recover $\dot{I}$ given the measurements $\dot{K}$.

For sparsely sampled MRI reconstruction, the k-space measurements $\dot{K}$ are undersampled for acceleration. The inverse problem for accelerated sparse sampling is thus ill-posed and typically solved as a regularized optimization problem with the objective of:

$$L = \left|\dot{M}\dot{F}\dot{S}\dot{I} - \dot{K}\right|_n + \dot{R} \quad (2)$$

Where $\left|\dot{M}\dot{F}\dot{S}\dot{I} - \dot{K}\right|_n$ is the fidelity term measuring the error between $\dot{M}\dot{F}\dot{S}\dot{I}$ and $\dot{K}$ and $\dot{R}$ is the regularization term characterizing the generic prior information. $\left|\dot{M}\dot{F}\dot{S}\dot{I} - \dot{K}\right|_n$ and $\dot{R}$ can be determined in various ways (e.g., total variation and wavelets constraint[33]) to meet the modeling assumption.

## 2.3 Fourier Series Representation

According to the Fourier basis approximation theorem, any square-integrable function on a finite interval can be approximated by a Fourier series, a linear combination of sines and cosines. Given a square-integrable periodic function $f(x)$ with the coordinate $x$ normalized to the interval of [0,1], its Fourier series representation is:

$$f(x) = a_0 + \sum_{n=1}^{\infty}(a_n \cos(2\pi nx) + b_n \sin(2\pi nx)) \tag{3}$$

Where the Fourier coefficients $a_n$ and $b_n$ are calculated by:

$$a_n = 2\int_0^T f(x)\cos(2\pi nx)\,dx \tag{4}$$

$$b_n = 2\int_0^T f(x)\sin(2\pi nx)\,dx \tag{5}$$

Equation (3) approximates $f(x)$ by decomposing it into sines and cosines of increasing frequencies, forming a Fourier basis bounded within [0,1].

### 2.4 Universal Approximation of INR with Sinusoidal Activation Functions

Considering an INR model with sinusoidal activation functions $g(x)$ with the representation of:

$$g(x) = W_2 \sin(W_1 x + b_1) + b_2 \tag{6}$$

Where $W_1$ and $W_2$ are weight matrix and $b_1$ and $b_2$ are bases. Due to the periodic nature of the sine function, the INR structure with such activations can approximate periodic signals.

According to the universal approximation theorem [34], MLPs with sufficient depth and width as well as appropriate activation functions, can approximate any continuous function. When using sinusoidal activations, INRs, formulated with MLPs, can represent a function in $L_2$ space (the space of square-integrable functions), as Fourier series do. This makes INRs well suited to approximate Fourier sequences directly. Mathematically, $g(x)$ can approximate any $f(x)$ expressible as a Fourier series:

$$f(x) \approx g(x) = \sum_{n=1}^{N} \alpha_n \sin(\omega_n x + \phi_n) \tag{7}$$

Where $\alpha_n$, $\omega_n$ and $\phi_n$ are parameters learned by $g(x)$ to best approximate $f(x)$, which matches the structure of Fourier decomposition. Therefore, while the depth and width are sufficiently large, a network can approximate functions that can be expressed as sums of sine and cosine functions, similar to a Fourier series.

### 2.5 k-space Representation in MRI

In MRI, k-space data represents the Fourier transform of the image in the spatial domain where each k-space point corresponds to a specific frequency component of the image. An INR with sinusoidal activations can directly learn the Fourier-space data by optimizing to fit sampled k-space points.

Given a sample set $\dot{K} = \{k_i\}$ for $i = 1, \dots, N$ fully sampled in k-space, with the corresponding coordinate $\dot{C} = \{c_i\}$ and complex signal value of $\dot{V} = \{v_i\}$ for $i = 1, \dots, N$. The INR model can be formulated as:

$$g: c \rightarrow v \quad with\ c \in [0,1], v \in \mathcal{C} \tag{8}$$

Where the input coordinate $c$ is normalized in k-space, and $\mathcal{C}$ represents the complex space. The network function $g$ maps k-space coordinates to the k-space complex signal value, which encodes the internal information of the entire k-space into the network parameters. $g$ is trained to minimize the difference between its prediction and the GT k-space intensity values, essentially learning the Fourier series coefficients. Let $\hat{g}(\dot{C})$ be the k-space representation predicted by the INR model. The training objective function can be formulated as follows:

$$L = \sum_{i=1}^{N} |\hat{g}(c_i) - f(c_i)|_n \tag{9}$$

Where $f(c_i)$ represents the GT at the sample point $c_i$. Through optimization in Equation (9), $g$ models the k-space signal as a Fourier series-like representation.

## 2.6 Training the Cohort Prior as a Regularized Approximation

Let the fully sampled GT k-space signal of a target patient be denoted as $f(\dot{C})$. The cohort that contains multiple patients is denoted as $\mathbb{C} = \{\dot{C}_i\}, i = 1, \dots, N$. Assuming that the cohort is well diversified and contains universal features of the target anatomy, training on a cohort of similar anatomical structures is to build a generalized model $g_{prior}(\mathbb{C}|\dot{\theta})$. $\dot{\theta}$ are the parameters of $g_{prior}$ representing the shared anatomical features observed within the cohort. Mathematically, this cohort-based model acts as a regularized approximation:

$$g_{prior}(\mathbb{C}|\dot{\theta}) \approx f(\mathbb{C}) \tag{10}$$

## 2.7 Sparse Data Consistency with Patient-specific Optimization

The new patient's sparse sampled MRI data can be denoted as $\dot{K}_s = \{k_{s,i}\}$, for $i = 1, \dots, M$ with corresponding k-space coordinate $\dot{C}_s = \{c_{s;i}\}$ and intensity of $\dot{V}_s = \{v_{s;i}\}$ for $i = 1, \dots, M$, where $M < N$. To optimize the prior for the specific patient, we aim to initialize the model with the cohort trained prior $\dot{\theta}$. Next, we adjust $\dot{\theta}$ from $g_{prior}(\mathbb{C}|\dot{\theta})$ to $g_{patient}(\dot{C}_s|\dot{\theta}_s)$ such that $g_{patient}(\dot{C}_s|\dot{\theta}_s)$ can accurately represent $f(\dot{C}_s)$ across the limited sample points:

$$L_{fidality\_sparse} = \sum_{i=1}^{M} |\hat{g}_{patient}(c_{s;i}|\theta_s) - f(c_{s;i})|_n \qquad (11)$$

Initializing from the cohort prior can be seen as a form of regularization to ensure the patient-specific fine-tuning does not overfit the artifact and noise in the sparse measurement but instead adheres to the common anatomical features captured from the cohort. This regularization can be expressed as a penalty term added to the loss:

$$L_{total\_sparse} = L_{fidality\_sparse} + \lambda R(g_{prior}(\mathbb{C}|\dot{\theta}), f(\dot{C})) \qquad (12)$$

Where $R(g_{prior}(\dot{C}|\dot{\theta}), f(\dot{C}))$ is the regularization term that penalizes deviation from the cohort-based prior. $\lambda$ is a hyperparameter specifying regularization strength resulting from the number of fine-tuning iterations and stopping threshold.

## 2.8 Large-scale Prior Training and Patient Specific Optimization

To better facilitate the generalizability of $g_{prior}$ across a large complex cohort, GAN [29] was applied during the training of the cohort prior $\dot{\theta}$. GAN-based training involves a generator $g$ and a discriminator $d$, where $g$ predicts the complex signals in k-space while $d$ differentiates between the GT k-space signal $f(\dot{C})$ and generated sampled $g(\dot{C})$.

The generator objective $L_g$ aims to minimize the likelihood that $d$ identifies $g(\dot{C})$ as fake and can be formulated as:

$$L_g = \mathbb{E}_{c \sim P_c}[\log(1 - d(g(\dot{C}|\dot{\theta})|\dot{\gamma}))] \qquad (13)$$

Where $\dot{\gamma}$ is the weight of discriminator $d$.

The discriminator objective $L_d$ aims to minimize its ability to classify real and fake samples and can be formulated as:

$$L_d = -\mathbb{E}_{f(c) \sim P_{real}}\left[\log\left(d(f(\dot{C})|\dot{\gamma})\right)\right] - \mathbb{E}_{c \sim P_c}[\log(1 - d(g(\dot{C}|\dot{\theta})|\dot{\gamma}))] \qquad (14)$$

The summation of $L_g$ and $L_d$ forms the fidelity terms for training $g_{prior}$. The total objective can be formulated as:

$$L_{total\_prior} = L_g + L_d \qquad (15)$$

At the patient-specific optimization stage, a straightforward iterative process, $ITR$, with $1 - SSIM$ fidelity term was employed to adjust the model towards new patient's sparse k-space measurements, where $SSIM$ refers to structure similarity indexed measurements and is fined in Equation (17). The cohort-trained prior was used as an implicit regularization term. Optimization proceeds until a predefined threshold (hyperparameter; elaborated in section 2.9 technical details) is met to avoid over-fitting noises and artifacts in the sparse measurements. $\lambda$ represents the regularization strength of $1 - SSIM$ fidelity term versus cohort-trained prior. $\lambda$ is inexplicitly defined by the number of optimizing iterations, where $N_{ITR} = 0$ defines $\lambda \to +\infty$ and $N_{ITR} \to +\infty$ defines $\lambda \to 0$. More details regarding the selection of the optimization threshold will be discussed in the following section (section 2.9 Technical details of k-INR). The objective for patient-specific optimization can be formulated as:

$$L_{total\_sparse} = \sum_{i=1}^{M}[1 - SSIM(\hat{g}_{patient}(c_{s;i}|\theta_s), f(c_{s;i}))] + \lambda R(g_{prior}(\mathbb{C}|\dot{\theta}), f(\dot{C})) \quad (16)$$

$$SSIM(g, f) = \frac{(2\mu_g\mu_f + c_1)(2\sigma_{gf} + c_2)}{(\mu_g^2 + \mu_f^2 + c_1)(\sigma_g^2 + \sigma_f^2 + c_2)} \quad (17)$$

Where $\mu_g$ and $\mu_f$ is the pixel mean of input $g$ and $f$ and $\sigma_{gf}$ is the covariance between $g$ and $f$, $\sigma_g^2$ and $\sigma_f^2$ is the variance of $g$ and $f$. Lastly, $c_1 = (k_1 L)^2$ and $c_2 = (k_2 L)^2$, where $k_1 = 0.01$ and $k_2 = 0.03$ in the current work and $L$ is the dynamic range of the pixel values ($2^{\#\ bits\ per\ pixel} - 1$).

## 2.9 Technical Details of k-INR

In our setup, we designed an MLP network with a periodic activation function [25] applied after each fully connected layer except the last output layer. For the UCSF STARVIBE liver dataset, a 22-layer MLP network with a width of 512 hidden neurons is selected. Since k-GINR trains and fine-tunes in three-dimensional (3D) k-space, the input feature to MLP was set to 3 to represent the 3D location $[k_x, k_y, k_z]$ of each sample point. The real and imaginary parts for each voxel intensity of each receiver coil (output) in k-space were modeled and predicted separately. Thus, the output feature was set to $2 \times nC$ with $nC$ represents the receiver coil number. As demonstrated by Figure 3, 1 - SSIM was chosen as the stopping criteria with a threshold (hyperparameter) tuned from the validation set at the fine-tuning stage (UCSF STARVIBE liver dataset: 0.13 for 3 times acceleration, 0.18 for 10 times acceleration and 0.21 for 20 times acceleration). The training and validation loss at the validation stage is defined in Equation (18-19) with $L_{train}$ penalizes on the dissimilarity between intermediate model prediction $\hat{g}_{patient}(c_{s;i}|\theta_s)$ and sparsely sampled k-space signals $f(c_{s;i})$ and $L_{val}$ penalizes on the dissimilarity between intermediate model prediction $\hat{g}_{patient}(c_{s;i}|\theta_s)$ and fully sampled k-space signals $f(c_i)$. For training the cohort prior, the Adam optimizer with a learning rate of $1e-6$ and epochs of 500 was used for optimization. For patient-specific fine-tuning, the Adam optimizer with a learning rate of $1e-5$ was used with maximum iterations of 2000. The network was implemented using PyTorch. For the forward model that projects the measurement signals to the spatial domain, the nonuniform Fast Fourier Transform (NUFFT) with coil sensitivity map generated slice-by-slice using ESPIRiT [35] with $20 \times 20$ calibration region was applied. Since ESPIRiT requires Cartesian k-space data, the raw radial k-space was first converted to image space via inverse NUFFT and then transformed back to Cartesian k-space using FFT. The forward model was implemented using the BART toolbox [36]. All the experiments were carried out on a 4×RTX A6000 GPU cluster with a batch size of $4 \times 2^{19}$, where $2^{19}$ is the number of voxel points.

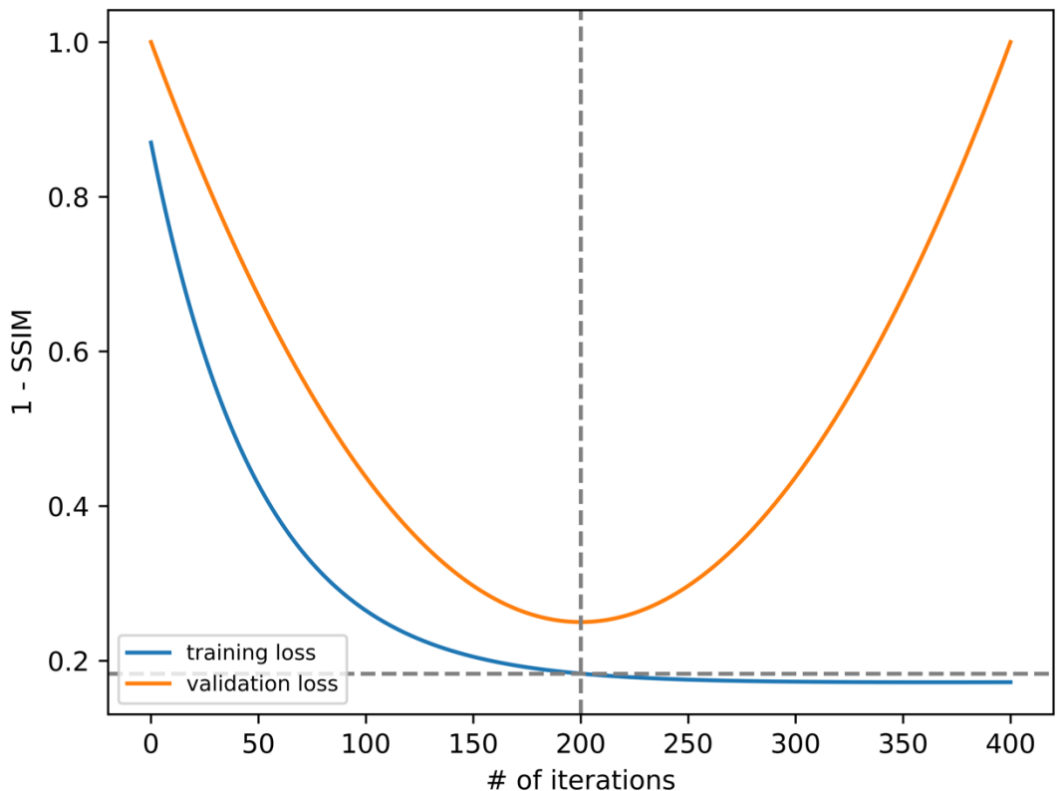

**Figure 3**: Demonstration of the hyperparameter tuning process on the validation set for the optimal stopping criteria for $1 - SSIM$ at the patient-specific fine-tuning stage. The current loss curve is demonstrated with the UCSF StarVIBE Liver dataset with 10x acceleration. The red star marks the optimal stopping point with the lowest validation loss.

$$L_{train} = \sum_{i=1}^{M} [1 - SSIM(\hat{g}_{patient}(c_{s;i}|\theta_s), f(c_{s;i}))] \qquad (18)$$

$$L_{val} = \sum_{i=1}^{M} [1 - SSIM(\hat{g}_{patient}(c_{s;i}|\theta_s), f(c_i))] \tag{19}$$

## 2.10    Baseline Algorithms and Model Evaluation

A conventional CS algorithm with total variation regularization [4] and an image-domain DL approach (3D Cascade CNN) [12] were included as baseline models. The CS algorithm was implemented with the *BART* package [36]. The 3D Cascade CNN method was implemented in house as the source code was not publically available.

The model performance was evaluated using the following metrics: root mean squared error (RMSE), peak-signal-to-noise ratio (PSNR), SSIM, and inference time, as shown in Equation (7, 20, 21).

$$RMSE = \sqrt{\frac{\sum_{i=1}^{N}(g(c_i) - f(c_i))^2}{N}} \tag{20}$$

$$PSNR = 20 \cdot \log_{10} \frac{MAX_I}{RMSE} \tag{21}$$

Where $MAX_I$ is the maximal possible pixel value in a tensor.

## 3. Experiments and Results

### 3.1 UCSF StarVIBE Liver Dataset

UCSF Liver StarVIBE was used as the second evaluation dataset. The study was approved by the local Institutional Review Board at UCSF (#14-15452) [13]. 118 patients were scanned on a 3T MRI scanner (MAGNETOM Vida, Siemens Healthcare, Erlangen, Germany) after injecting hepatobiliary contrast (gadoxetic acid; Eovist, Bayer). A prototype free-breathing T1-weighted volumetric golden angle stack-of-stars sequence was used for MRI signal acquisition. The scanning parameters were - TE=1.5 ms, TR=3 ms, matrix size ($nh \times nw$) = 288 x 288, field of view (FOV) = 374 mm x 374 mm, in-plane resolution = 1.3 mm × 1.3 mm, slice thickness = 3 mm, RV per partition ($nViews$) = 3000, $nC$ = 26, $nz$ = 64-75, acquisition time = 8-10 min. K-space data was sorted into 8 motion bins to reconstruct motion resolved 4D MRI [37], [38]. To accommodate available computational resources without losing generality, we only kept one motion bin with the most number of spokes of each sequence (in the range of [427, 465]) and treated it as the fully sampled sequence (Based on Equation (19), fully sampled radial images require $\geq nx \times \frac{\pi}{2}$ spokes, resulting in $\geq 288 \times \frac{\pi}{2} = 452$ spokes for a matrix size of $288 \times 288$, where $nx$ is the matrix size in one dimension.).

$$nViews \approx \frac{\pi}{2} \times nx \qquad (22)$$

To effectively manage the volume of training input, only 40 continuous axial slices that contain liver morphologies were included ($nz' = 40$). The selected axial slices were manually determined for each patient. Thus, the resultant k-data is of shape $nx \times nViews \times nz' \times nC = 288 \times 500 \times 40 \times 26$. The training, validation and testing patient split is $82:18:18 \approx 7:1.5:1.5$.

For validation and testing patients, retrospective under-sampling was performed by keeping the first $ceil(\frac{1}{3})$, $ceil(\frac{1}{10})$, and $ceil(\frac{1}{20})$ spokes of the fully sampled spokes, respectively, corresponding to acceleration rates of 3x, 10x, and 20x.

## 3.2 Experiments

Figure 4 illustrates the trained prior (first row) in the k-space (showing the real and imaginary components of the first coil, along with their separate heatmap distributions) and the spatial domains, respectively. A GT patient example from the training cohort (second row) is provided for comparison. In k-space, we observe that the pixel distribution in the peripheral high-frequency region of the training prior (third row) is less dense than that of the GT, with the difference heatmap in k-space (third column) highlighting more pronounced differences in the high-frequency regions. The cohort prior exhibits less detailed anatomical structures, focusing more on general features in comparison to the patient-specific signal distribution.

Table 1 and Figures 5, 6, and 7 present the statistical and visual results for the test cohort using 3x, 10x, and 20x accelerated acquisitions, respectively. Figure 5 demonstrates that both k-GINR and CS deliver robust reconstruction at 3x and 10x accelerations, with k-GINR showing moderately better artifact reduction and detail preservation. At 20x acceleration, CS exhibits visible detail loss and pronounced streaking artifacts, whereas k-GINR maintains performance with only a moderate increase in noise. In contrast, Cascade CNN produces overly smoothed reconstructions at all acceleration levels, with severe loss of anatomical detail as acceleration increases.

In the line profile comparison shown in Figure 5, Cascade CNN profiles deviate notably from the GT in both amplitude and patterns, indicating unacceptable degradation in contrast and anatomical fidelity. Figure 6 highlights a multi-slice axial comparison of liver patient reconstructions between k-GINR and Cascade CNN. k-GINR preserves accurate anatomical details of the liver tumor across all acceleration levels, while Cascade CNN results become unusable beyond 3x acceleration due to loss of imaging details and hallucinated structures such as non-existent liver mass.

The proposed k-GINR consistently demonstrates robust reconstruction of raw coil data for liver patients at all acceleration levels, with more pronounced advantages at higher acceleration rates. In terms of speed, k-GINR achieves an average optimization time of approximately 197 seconds, slightly faster than CS (245 seconds) but significantly slower than Cascade CNN, which completes reconstruction in sub-second, without requiring test-phase optimization.

Figure 8 illustrates the patient-specific optimization process during the test phase, starting with weights pre-trained on the training cohort prior and using 10x accelerated k-space signals as input. The left-hand side of Figure 8 represents more inexplicit regularization with prior from the training cohort (fewer optimization iterations). In comparison, the right-hand side of Figure 8 represents more inexplicit regularization with accelerated k-space signals (more optimization iterations). The red star marks the optimal stopping point of 1-SSIM determined through hyperparameter tuning on the validation set. This point represents the ideal balance between training the prior and reconstructing the accelerated k-space signal, effectively preserving more anatomical details while minimizing artifacts caused by acceleration.

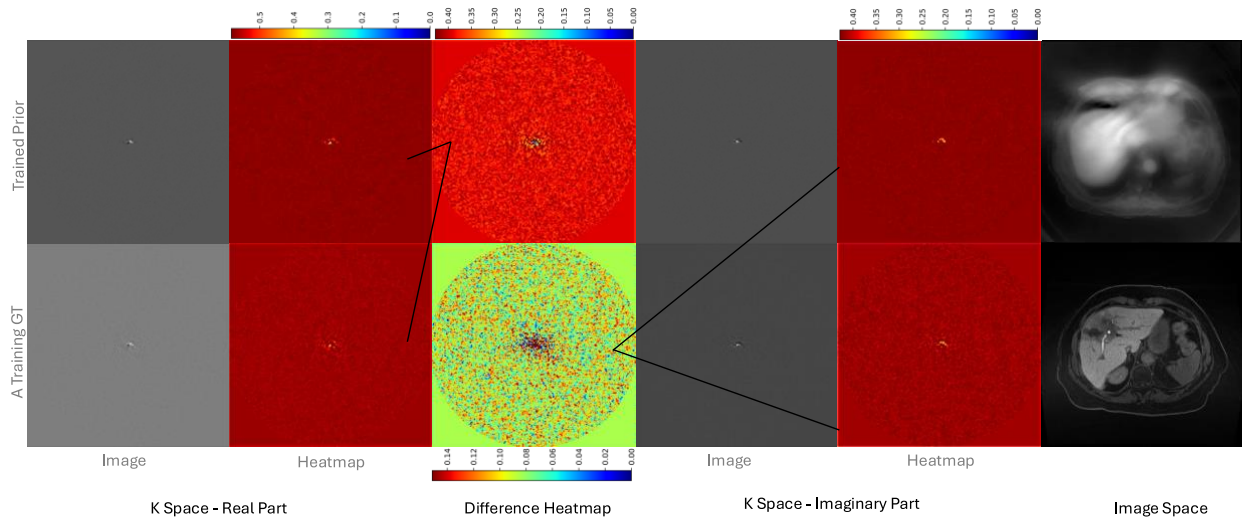

**Figure 4**: K space (first, real component, and fourth, imaginary component, columns) and image domain (sixth column) visualization of the trained prior of the UCSF STARVIBE liver dataset compared with a patient instance in the corresponding training cohort. The heatmap of k-space visualization (second and fifth columns) and the absolute difference heatmap (third column) between the patient instance and its corresponding trained prior in k-space are also presented. All the images are normalized to the scale of [0,1] for visualization.

|  | Acceleration | SSIM↑ | RMSE↓ | PSNR (dB) ↑ | Averaged Time (s) ↓ |
|---|---|---|---|---|---|
| **k-GINR** | 3x | **0.93 ± 0.02** | **0.03 ± 0.01** | **29.27 ± 2.13** |  |
|  | 10x | **0.89 ± 0.04** | **0.05 ± 0.03** | **26.75 ± 2.89** | 197 |
|  | 20x | **0.85 ± 0.07** | **0.08 ± 0.07** | **22.43 ± 3.21** |  |
| **Cascade CNN** | 3x | 0.85 ± 0.06 | 0.08 ± 0.09 | 22.15 ± 2.98 |  |
|  | 10x | 0.79 ± 0.08 | 0.13 ± 0.11 | 17.33 ± 3.27 | **<1** |
|  | 20x | 0.67 ± 0.13 | 0.25 ± 0.15 | 12.08 ± 4.89 |  |
| **Compressed Sensing** | 3x | 0.91 ± 0.02 | 0.05 ± 0.02 | 25.28 ± 2.65 |  |
|  | 10x | 0.86 ± 0.05 | 0.11 ± 0.07 | 19.21 ± 3.01 | 245 |
|  | 20x | 0.74 ± 0.11 | 0.17 ± 0.12 | 15.36 ± 4.29 |  |

**Table 1**: Quantitative evaluation results of the UCSF StarVIBE liver dataset in the image domain of the test cohort using the proposed k-GINR and benchmark Cascade CNN as well as CS models. Three acceleration ratios, including 3 times acceleration (3x), 10 times acceleration (10x), and 20 times acceleration (20x), are presented. Structural indexed similarity measurement (SSIM), root

mean squared error (RMSE), peak signal-to-noise ratio (PSNR), and averaged inference time are reported. The arrow after each metric indicates the desired direction for a better reconstruction result. The results from the best performers are bolded. All the images are normalized to the scale [0,1].

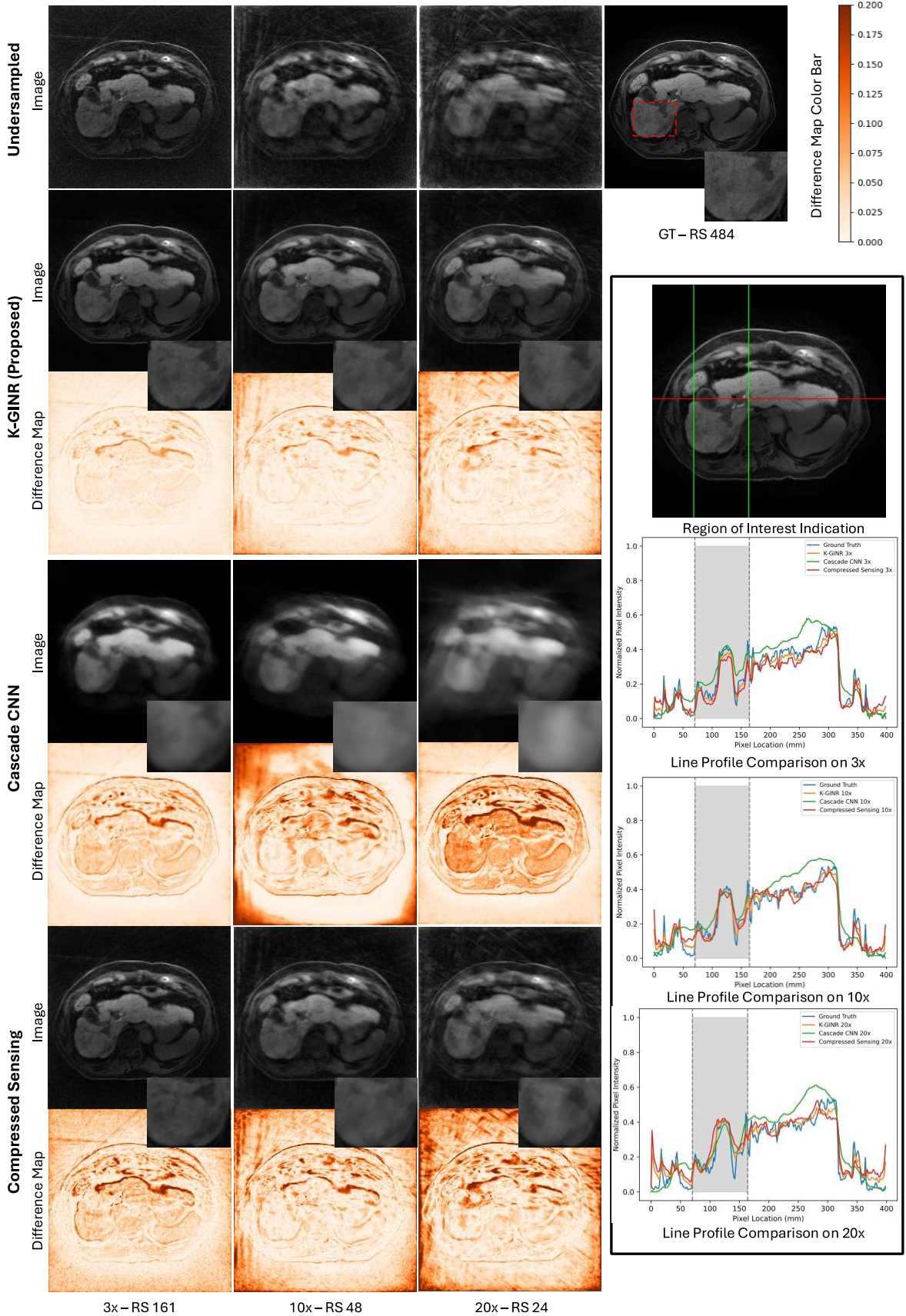

**Figure 5**: Image domain visualization of a selected patient in the test cohort of the UCSF STARVIBE liver dataset. Reconstructions from our proposed K-GINR (2$^{nd}$-3$^{rd}$ rows) and the benchmark cascade CNN (4$^{th}$-5$^{th}$ rows) and CS (6$^{th}$-7$^{th}$ rows), along with its residual maps in comparison with GT are presented. The color bar for the residual maps is visualized in the bottom right corner. From left to right, the reconstruction results from 5 times acceleration/radial spoke 40 (5x – RS 40), 10 times acceleration/radial spoke 20 (10X - RS 20), and 20 times acceleration/radial spoke 10 (20X - RS 10) are visualized in the first, second, and third columns. A zoomed-in detail visualization is attached to the GT, k-GINR, Cascade CNN, and CS reconstructions with a region of interest defined in the GT image. Pixel intensity horizontal line profile distributions (indicated with red line) reconstructed by the proposed and benchmark algorithms across different acceleration ratios are shown in the bottom right corner. All the images are normalized to the scale [0,1] for visualization.

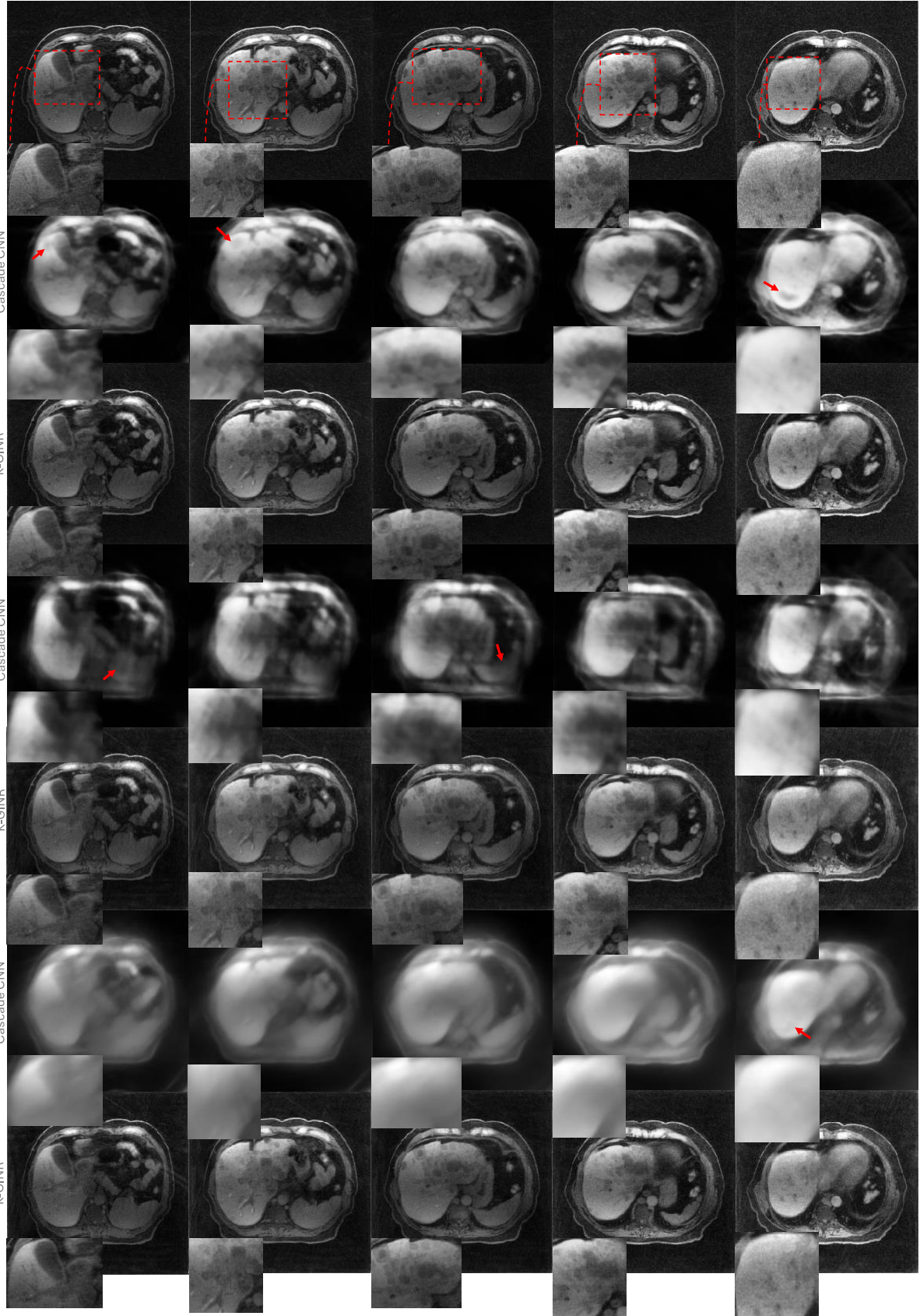

**Figure 6**: Image domain visualization of a selected patient in the test cohort of the UCSF STARVIBE liver dataset. Predictions results of k-GINR and Cascade CNN across different axial slices at 3x (2-3 rows), 10x (4-5 rows), and 20x (6-7 rows) acceleration ratios are presented. The red boxes in the GT images show the region of interest selected for zoom-in view at each axial slice. The red arrows mark the artifacts presented in the Cascade CNN reconstructions.

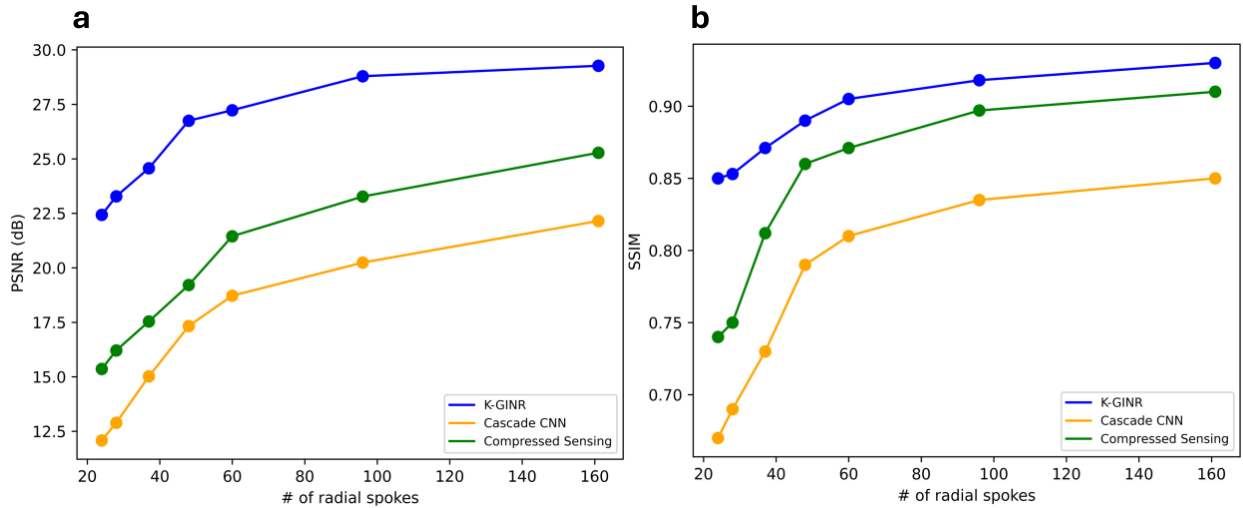

**Figure 7**: The k-GINR and selected benchmark reconstruction results (PSNR and SSIM) referencing different numbers of sampled radial spokes at the test phase of the UCSF STARVIBE liver dataset.

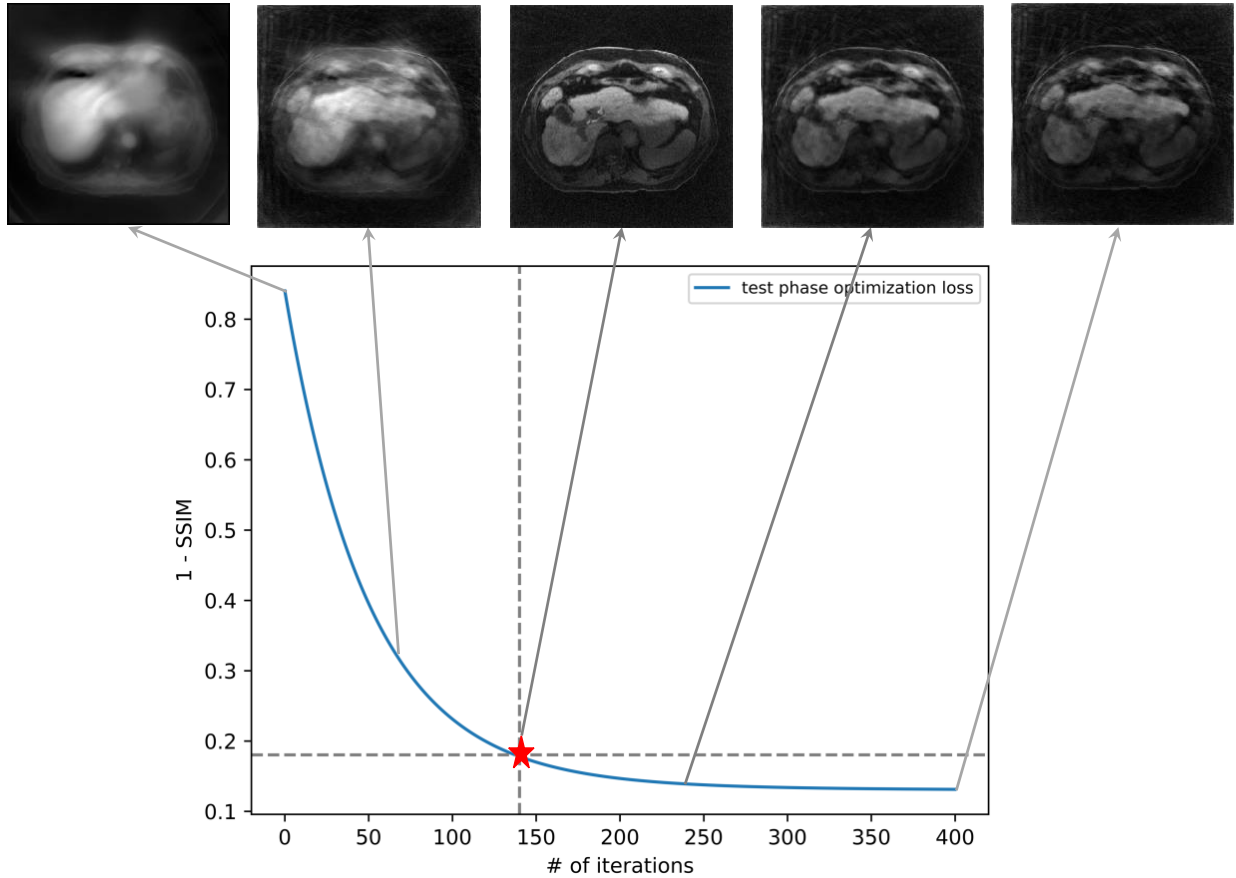

**Figure 8**: The patient-specific optimization process at the test phase demonstrated with the UCSF StarVIBE Liver dataset at a 10x acceleration ratio. The bottom row shows the loss corresponding to the optimizing iterations. The top rows show prediction results sampled from K-GINR at different optimizing iterations. The red star marks the optimal loss-stopping threshold determined via the hyperparameter tuning step (Figure 3).

## 4. Discussion

Although numerous DL models have shown significant promise for accelerated MRI reconstruction over the past decade [39], four notable challenges are intrinsic to the network architecture. First, DL models, especially architectures relying on convolutional discrete filters, often struggle with reconstructing high-frequency components in k-space, leading to overly smoothed reconstruction and a loss of anatomical details. Second, DL models can be prone to hallucination, particularly in image-domain-based methods with high undersampling ratios (example demonstrated in Figure 8). Such hallucinations can mislead diagnosis, treatment planning, and subsequent therapy delivery [40], [41]. Third, the existing DL methods train their models on a large cohort to establish prior and then inference using the incoming patient data as input without patient-specific modeling tuning. Such a framework can be limited in adapting to individual patient-specific anatomies for more accurate reconstruction [42]. Fourth, most current DL models are incompatible with non-Cartesian k-space trajectories widely used in fast or motion-robust MR acquisitions [43], [44]. Compared with Cartesian sequences, non-Cartesian sequences are more resilient to aliasing and motion artifacts [45].

The proposed k-GINR addresses these challenges with novel reconstruction architecture and formulation. First, the proposed k-GINR model leverages an INR framework [25] that inherently preserves high-frequency information with its continuous representation throughout the reconstruction process. This allows k-GINR to better maintain fine-scale k-space details crucial for accurate boundary definition. Second, k-GINR reduces hallucinations by incorporating raw k-space signals without additional processing of the sparse sampling, minimizing the generation of spurious features. The GAN-based training [29] further helps the network learn common patterns from the GT data. Third, k-GINR supports patient-specific adaptation through a modified two-stage approach: 1) initial training on a large cohort to generate a universal prior (fully supervised), and 2) fine-tuning on patient-specific sparse data for further optimization (self-supervised). This additional self-supervised fine-tuning step significantly improves the balance between statistical learning and individual patient characteristics, as demonstrated by our brain and liver reconstruction results. Lastly, using MLP in k-GINR overcomes the limitations of CNNs and transformers with non-Cartesian k-space trajectories.

k-GINR differs from the pioneering study by Shen et al. [28], which introduced an INR-based NERP framework for reconstructing sparse non-Cartesian k-space data using priors learned from the same patients' previous scans. The effectiveness of NERP thus relies on the availability of images from the same patient, technique, and anatomy. The stringent condition limits its broader applicability. In cases where longitudinal images are unavailable or significant anatomical changes occur between scans, NERP cannot be effectively trained. In contrast, k-GINR significantly relaxes the condition and only requires populational patient data for training.

Given its enhanced speed and precision from CS, the integration of k-GINR into clinical MRI workflows holds significant promise for enhancing imaging speed and precision, particularly in applications requiring high-resolution, motion-resolved imaging and MRI-guided interventions.

High-resolution imaging is essential for abdominal imaging, including the liver and pancreas, where accurate delineation of anatomical structures and detection of small pathological changes are crucial. For instance, high-resolution MRI enables early detection of liver fibrosis and hepatocellular carcinoma [46]. Similarly, MR cholangiopancreatography improves the visualization of ductal abnormalities and small neoplastic lesions in pancreatic tumors [47], [48]. High-resolution imaging with precise tumor boundary contouring plays a pivotal role in enabling advanced interventions, such as stereotactic body radiation therapy, for hepatocellular carcinoma or liver metastases, where conformal and high-dose radiation is delivered to the tumor while avoiding excessive healthy tissue toxicity [13], [40]. Abdominal MR is particularly affected by respiratory and digestive motion, which, if unmanaged, leads to severe artifacts and unusable images. MR performed in a single breath hold reduces the impact of motion artifacts but is limited by how long the patient can hold the breath. It often requires aggressive acceleration that leads to compromised imaging quality. Combining images acquired during multiple breath-holds improves the signal but unavoidably prolongs acquisition and are subject to registration error [49], [50]. k-GINR's potential to reconstruct aggressively undersampled non-Cartesian sequences while better preserving fine structural details can be immensely valuable here.

Besides static instances, motion-resolved imaging is crucial for capturing dynamic physiological processes in applications, including cardiac MRI. Non-Cartesian sampling trajectories are resilient to motion artifacts, such as the ones induced by respiratory or cardiac movements [45]. 4D MR sequences are particularly beneficial for radiotherapy, where tumor contours in individual breathing phases are used to determine the internal target volume (ITV). Acquiring such images requires sampling individual respiratory phases, which can be lengthy, yet the full sampling condition may not be met for irregular breathers[51], leading to motion artifacts that affect the ITV definition. In other words, the varying sampling rates in motion-resolved MR can lead to worse undersampling conditions for certain breathing phases than the average sampling rate suggests. The problem is exacerbated by accelerated acquisition. Therefore, k-GINR, showing greater resilience to more aggressive undersampling conditions, can improve the robustness and acquisition efficiency of 4DMRI of moving anatomies. This acceleration allows clinicians to visualize rapid physiological motions, supporting accurate functional analysis and treatment planning with enhanced efficiency [52].

Our proposed method is not without room for improvement. First, the current study of k-GINR is limited to a resolution of $nx \times nViews \times nz' \times nC = 288 \times 500 \times 40 \times 26$ to fit in the 4×RTX A6000 GPU memory. Achieving higher-resolution reconstructions requires a more complex perceptron design and correspondingly larger GPU memory. While rapid advances in GPU technology and memory capacity, benefitting from the explosive interest in the large language models, will likely alleviate this constraint, developing more GPU-efficient architectures to handle higher resolutions and larger imaging volumes will be valuable. Second, the current inference time of k-GINR (up to around 3 minutes) may be sub-optimal for online applications. Speeding up the inference stage can be achieved through strategies such as network pruning—reducing less critical weights or neurons while preserving accuracy [53] and partial k-space optimization - focusing on optimizing the high-frequency (outer k-space) components to enhance detailed reconstructions.

## 5. Conclusion

k-GINR is a novel INR network with adversarial training designed for direct undersampled k-space reconstruction. It outperforms a DL method trained in the image domain and a conventional compressing method, demonstrating superior reconstruction results on the raw UCSF StarVIBE liver radial k-space data for a wide range of undersampling ratios. K-GINR is uniquely suited to reconstruct aggressively undersampled non-Cartesian MR.